\documentstyle[twocolumn,prl,aps,floats]{revtex}
\begin{document}
\input{psfig.sty}
\draft

\wideabs{
\title{UCN upscattering rates in a molecular deuterium crystal}

\author{C.-Y. Liu$^1$, A.R. Young$^1$, and S. K. Lamoreaux$^2$}

\address{1.  Princeton University, Physics Dept., Princeton, NJ 08544\\
2. University of California, Los Alamos National Laboratory,
Physics Division P-23, Los Alamos, NM 87545}

\date{\today}

\maketitle

\begin{abstract}

A calculation of ultra-cold neutron (UCN) upscattering rates in 
molecular deuterium solids has 
been carried out, taking into account intra-molecular excitations
and phonons.  The different molecular species 
ortho-$D_2$ (with even rotational quantum number J) and para-D$_{2}$ (with odd J)
exhibit significantly different UCN-phonon annihilation cross-sections.
Para- to ortho-D$_{2}$ conversion, furthermore, couples UCN to an
energy bath of excited rotational states without mediating phonons.
This anomalous upscattering mechanism restricts the UCN lifetime to
4.6 msec in a normal-D$_{2}$ solid with 33\% para content.

\end{abstract}

\pacs{25.40.Fq, 29.25.Dz}}

The low density of ultra-cold neutrons (UCN) available using conventional
cold moderators at nuclear reactors has long been the main constraint 
in the pursuit of high precision measurements of neutron $\beta$-decay with
UCN\cite{beta_decay}.
To pursue the possibility of utilizing this simple hadronic system for tests of weak interaction theories,
Golub and Pendlebury~\cite{superthermal} proposed a way to increase 
UCN production 
through the exchange of energy between a cold neutron bath and
the phonons in certain cold moderators 
such as super-fluid $^4$He, with large neutron scattering cross-sections and 
small neutron absorption cross-sections. 
Superthermal UCN sources exploit the characteristics of low temperature
substances, in which a large number of phonon modes are available for neutron
downscattering, while the number of phonons present which can upscatter UCN
is suppressed.  Ideally, the density of UCN produced in such a source is 
limited either by nuclear absorption in the moderator as is the case for
solid deuterium, or the neutron lifetime itself as is the case for
superfluid $^{4}$He.

A solid D$_{2}$ UCN superthermal source~\cite{UCNbook}
is under development at the Los Alamos Neutron Science Center. 
Preliminary estimates~\cite{Malik&Golub} promise gains 
in the available UCN density
as high as two orders of magnitude
over existing UCN facilities. To characterize its performance,
calculations of essential physical parameters,
such as scattering cross-sections, UCN residence time, etc., with more
detailed and accurate models than those presently available are in
increasing demand. 
In this report, we demonstrate that in the presence of even very small
concentrations of para-D$_{2}$ (with total nuclear spin 1)
can 
dominate the UCN upscattering rate, overwhelming the usual phonon
annihilation mechanism. This results in greatly reduced UCN lifetimes in
the solid and orders of magnitude reductions in the
achievable UCN density.

The deuterium molecule is a two-body system with the quantum properties of 
identical bosons. Its nuclear spin wave-function couples to molecular
rotational states with same parity to preserve the symmetry of the
wave-function under permutation of identical particles. Koppel and 
Young~\cite{Koppel&Young}
calculated the neutron scattering 
cross-section of this molecular system, taking into consideration 
induced transitions between molecular rotational and vibrational states.
In their formulation, an incoherent approximation was used and  
translational coordinates were assumed to commute with inter-molecular
degrees of freedom. The derived double differential cross-section in which 
the interference term is neglected has the form, for UCN with incident 
wavenumber $k$ and final wavenumber $k'$,
\begin{eqnarray}
\ && \frac{d^{2}\sigma} {d\Omega d\epsilon} = \frac{1}{2\pi\hbar} \frac{k'}{k} 
    \int_{-\infty}^{\infty} dt \nonumber \\
\ && \sum_{l} <\phi_{it}|e^{-i\kappa \cdot r_{l}(0)}
    e^{i\kappa \cdot r_{l}(t)}|\phi_{it}>_{Trans} \nonumber \\
\ &\times& \sum_{J} P_{JS} 
   \sum_{J'} {\cal S}_{JJ'} (2J' +1)e^{i(E_{J'}-E_{J})t/\hbar} \nonumber \\
\ &\times& \sum_{n=0}\frac{e^{in\omega t}}{n!}(\frac{\hbar^{2}\kappa^{2}}{2M_{D_{2}}\hbar\omega})^{n}
    \sum_{l=|J'-J|}^{J'+J}|A_{nl}|^{2}C^{2}(JJ'l;00),
\label{double_cross_section}
\end{eqnarray} 

\noindent where $\hbar \kappa$ is the momentum transfer of the scattered neutron,
$P_{JS}$ is the population of the initial molecular state with a total
nuclear spin $S$ and rotational quantum number $J$,
$E_{J}=7 \textrm{meV} \times J(J+1)/2$ is the rotational spectrum, $\hbar \omega$ is the 
inter-molecular vibrational energy with
$n$ characterizing the number of vibrational energy quanta, and $C(JJ'l;00)$ 
is a Clebsh-Gordon coefficient. $A_{nl}$ is defined as an integral over the 
orientation of a molecule, i.e.,
\begin{equation}
\ A_{nl} = \int_{-1}^{1} d\mu \mu^{n} exp(-\frac{\hbar\kappa^{2}\mu^{2}}{4M_{D_{2}}\omega}+\frac{i\kappa a \mu}{2})P_{l}(\mu),
\end{equation}

\noindent with $a = 0.74 \AA$ the equilibrium separation distance of the D-D bond,
$\mu$ the cosine of the inclination angle of the molecular axis
from the z axis of a reference Euclidean coordinate system, and $P_{l}$
the Legendre polynomial of order $l$.  

The input parameter ${\cal S}_{JJ'}$ (in units of barns)
in (\ref{double_cross_section}) for transitions between different
rotational states is deduced and listed in table~\ref{scattering_length}. 
Note here that only the incoherent scattering length $a_{inc}$ of a bound
nuclide contributes to the ortho(even $J$)/para(odd $J$) conversion.

\begin{table}

\caption{Intrinsic scattering cross-sections  ${\cal S}_{JJ'}$ 
associated with  different rotational transitions [6]. }

\label{scattering_length}

\begin{tabular}{ccc}
$\cal{S}_{JJ'}$ & Even-J(Ortho) & Odd-J(Para)  \\ \hline  
Even-J$^{'}$ & $a_{coh}^{2}+\frac{5}{8}a_{inc}^{2}=6.687/4\pi$ 
& $\frac{3}{4} a_{inc}^{2}=1.530/4\pi$ \\
Odd-J$^{'}$ & $\frac{3}{8}a_{inc}^{2}=0.765/4\pi$ & 
$a_{coh}^{2}+\frac{1}{4}a_{inc}^{2}=6.102/4\pi$ \\ \hline 
\end{tabular}
\end{table}

In a D$_{2}$ solid, the populations of the even-$J$(ortho state) and
the odd-$J$(para state) are typically determined by the ortho/para
population of the gas phase before the D$_{2}$ is frozen into the solid. 
The self-conversion between these two species in the solid phase into
a thermal Boltzman distribution is extremely  
slow compared with the time scale of experiment. For example, a room
temperature equilibrium D$_{2}$ is a mixture of 67\% ortho-D$_{2}$ and 33\%
para-D$_{2}$. The conversion rate in the solid is measured to be
0.06\%/hr~\cite{silvera}, requiring about 7 months to reduce the
para content to 1.65\% from 33\%. 
On the other hand, relaxation to thermal distributions is rapid among 
ortho- and para- species themselves. Consequently,  
in low temperature circumstances relevant to super-thermal solid deuterium
source(T$<$20K), only ground states($J=0$ for ortho; $J=1$ for para)
are present.

In the case of neutrons scattered off a low temperature crystal with well-defined lattice
structures, harmonic solid correlation functions should be applied to the translational
part of (\ref{double_cross_section}). Following the standard treatment of 
lattice dynamics~\cite{Lovesey}, we perform a phonon expansion 
\begin{eqnarray*}
\ <&exp&\{-i\kappa \cdot r_{l}\}exp\{i\kappa \cdot r_{l'}(t)\}>_{Trans} \\
\ &=& exp\{-2W(\kappa)\}exp\{<\kappa \cdot u_{l} \kappa \cdot u_{l'}(t)>\}, \\
\ &=& exp\{-2W(\kappa)\}[ 1 + <\kappa \cdot u_{l} \kappa \cdot u_{l'}(t)> \\
\ & & +O(<\kappa \cdot u_{l} \kappa \cdot u_{l'}(t)>^{2}) ].
\end{eqnarray*}

\noindent An overall Debye-Waller factor is extracted in front, and only the
first two terms are left for discussion, yielding the zero and one
phonon exchange processes, respectively. 

Unlike conventional applications of elastic solid correlation functions,
the first term in the phonon expansion(zero phonon term) coupled to molecular
internal energy states not only gives rise to UCN energy transition(mainly upscattering), but overwhelms
any phonon contributions when para-D$_{2}$ is present. The conversion of para- into ortho- molecules provides
direct energy transfer to UCN.
This scattering 
cross-section without phonon couplings has the simple form:
\begin{eqnarray} \label{upscatter}
\ (\frac{d\sigma}{d\Omega})_{J=1\rightarrow 0}^{0\mbox{ }phonon} &=& \frac{3}{4}a_{inc}^{2}\frac{k'}{k}
e^{-2W(\kappa)}\\ \nonumber
\ && \times \left[ 4j_{1}^{2}(\frac{\kappa a}{2}) C^{2}(101;00)\right],  
\end{eqnarray}
 
\noindent where $j_1(\frac{\kappa a}{2})$ is a spherical Bessel function of 
order one\cite{Temme}.  The increase
of the neutron momentum is definite, i.e.,
\begin{equation}
\  k'=\sqrt{2m_{n}\Delta E_{10}}/\hbar.
\label{momentum_transfer}
\end{equation}
  
\noindent A conversion energy $\Delta E_{10}$ of 7 meV gives $k'$ a value of 1.84$\times$10$^{10}$m$^{-1}$.
The momentum transfer $\kappa$ can be well-approximated by $k'$ for
UCN ($k'>>k_{ucn}$=1.27$\times$ 10$^{8}$m$^{-1}$), and the 
above differential cross-section is isotropic, making
the integration of (\ref{upscatter}) straightforward. 
A Debye-Waller factor originating from the uncertainty of positions of lattice 
sites, reduces the amplitude by a factor of 0.76. 
The temperature independent\footnote{The only temperature dependence comes
from a roughly 5\% decrease in the Debye-Waller factor between 4 and 18K.}
total cross section of UCN upscattering $\sigma_{10}$ is calculated to be 31 barns. This is at least 
of an order of 
magnitude larger than the phonon annihilation cross-section in a 4K solid.  

The rate of loss of UCN in the solid is
\begin{eqnarray}
\ \dot{\rho}_{ucn} &=& w_{fi}/V, \\ \nonumber
\                  &=& \rho_{ucn} [\sigma_{10}v_{ucn} \rho'].  
\end{eqnarray}
 
\noindent Here $\rho'$ is the density of para-D$_{2}$, taken to be
3$\times$10$^{22}$ cm$^{-3}$. The corresponding upscattering time $\tau_{up}$
of UCN in a pure para-D$_{2}$ molecular solid
is therefore
\begin{eqnarray*}
\ \tau_{up} &=& \frac{1}{\sigma_{10} v_{ucn} \rho'}, \\
\                     &=& 1.5 \qquad \textrm{msec}.
\end{eqnarray*}

\noindent For normal-D$_{2}$ which retains the room temperature equilibrium ortho/para 
ratio, the upscattering time due to the spin relaxation of 
para species is 4.6 msec!

\begin{figure}
\centerline{\psfig{figure=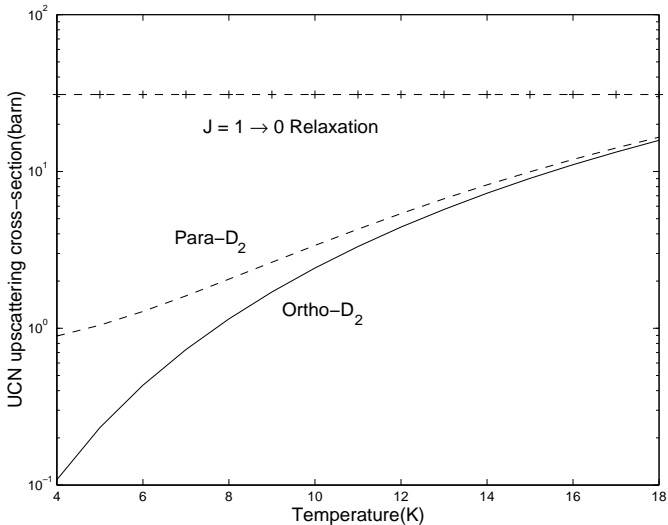,width=3.5 in}}
\caption{UCN upscattering cross-section vs. temperature of solid D$_{2}$.  
The one-phonon annihilation cross-section 
in an ortho-D$_{2}$ solid (solid curve)
and in a para-D$_{2}$ solid(dashed curve) are plotted. The dashed-dagger
line is the temperature independent UCN upscattering cross-section involving J=1$\rightarrow$0 relaxation not coupled to phonons
in a para-D$_{2}$ solid.}
\label{figu}
\end{figure}
  
To estimate the phonon upscattering rates, we approximate the SD$_{2}$ 
hcp/fcc lattice as a cubic lattice to simplify the treatment of polarization
anisotropies. 
The expression of the incoherent double differential cross-section involving 
one phonon exchange is 
\begin{eqnarray}
\ &(&\frac{d^{2}\sigma}{d\Omega dE^{'}})^{1\mbox{ } phonon}_{J \rightarrow J'} = 
\frac{k^{'}}{k} \frac{\hbar^{2}\kappa^{2}}{2M_{D_{2}}} e^{-2W(\kappa)}
  {\cal S}_{JJ'} (2J'+1) \nonumber \\
\ &\times& \sum_{n}
(\frac{\hbar\kappa^{2}}{2M_{D_{2}}\omega})^{n}\frac{1}{n!} 
\sum_{l=|J'-J|}^{J'+J} |A_{nl}|^{2} C^{2}(JJ'l;00) \nonumber \\
\ &\times& \frac{Z(E_{ph})}{E_{ph}} 
 \left\{ \begin{array}{ll}
         n(E_{ph})+1 & \textrm{if E$_{ph} >= 0$ } \\
         n(E_{ph})   & \textrm{if E$_{ph} < 0$,}
         \end{array} \right.
\label{one_phonon}
\end{eqnarray}

\noindent where the energy of phonon $E_{ph} = \epsilon +
\Delta E(J'\rightarrow J) + n\hbar\omega$, complying with the law of conservation
of energy. Positive and negative values of $E_{ph}$ correspond to 
single phonon creation and annihilation, respectively. $Z(E)$ represents
the normalized phonon density of states, and $n(E)$ the occupation number
of phonons with energy $E$. 

With a simple Debye model, in which 
\begin{equation}
\ Z(E)=\frac{3E^{2}}{(k_{B}T_{\Theta})^{3}} ,
\end{equation}
\noindent and the Debye temperature $T_{\Theta}$(110K for D2) is the 
only parameter, a double integration of (\ref{one_phonon}) with
initial energy of UCN (see Figure \ref{figu}) reproduces the upscattering cross-sections 
in ortho-deuterium calculated
by Yu, Malik and Golub~\cite{Malik&Golub}, in whose treatment rotational 
transitions were not considered. 
In an ortho-D$_{2}$  
solid, the $J=0\rightarrow 0$ process dominates the upscattering. 
Even though the $J=0 \rightarrow 1$ transition is energetically allowed
through coupling to a phonon, the cross-section is kinematically suppressed by
the smaller final phase space of upscattered neutrons.  
Para-D$_{2}$ has a distinguishably larger one-phonon annihilation cross-section than the ortho species.  
The origin of difference is again related to the $J=1\rightarrow 0$ relaxation channel.
This provides UCN with additional energy to scatter into a larger volume 
of phase space,
and secondly, it couples UCN to high energy phonons with large density of states, and is 
thus less restricted by the availability of phonon modes than
the $J=1 \rightarrow 1$ process. 
However, it is still suppressed by its small coupling to the phonon field, as opposed to the
zero phonon term.

In summary, para-deuterium has a spin relaxation channel in which its conversion
energy of 7 meV can be released to UCN, resulting in a temperature-indepedent short UCN lifetime of 4.6 msec in
a normal-D$_{2}$ solid.  Elimination of the para-D$_{2}$ is necessary
to achieve UCN lifetimes comparable to the nuclear absorption time in solid
D$_{2}$.

The authors wish to acknowledge the generous support of the National Science
Foundation through grant NSF-9807133
and the Department of Energy through LDRD-8K23-XAKJ funds.

\end{document}